\documentclass[aps,prb,amsmath,amssymb,amsfonts,superscriptaddress,floatfix]{revtex4} 
\pdfoutput=1
\usepackage{amsmath,amsthm,amsfonts,amssymb}
\usepackage{graphicx}

\usepackage{color}
\numberwithin{equation}{section}
\numberwithin{figure}{section}

\newtheorem{theorem}{Theorem}
\newtheorem{lemma}{Lemma}

\newtheorem{definition}{Definition}
\newtheorem{conjecture}{Conjecture}
\newtheorem{corollary}{Corollary}

\DeclareMathAlphabet{\mathpzc}{OT1}{pzc}{m}{it}

\newcommand{\be}{\begin{equation}}
\newcommand{\ee}{\end{equation}}

\newcommand{\mZ}{{\mathbb Z}}
\newcommand{\mF}{{\mathbb F}}
\newcommand{\vol}{{\rm vol}}
\newcommand{\mC}{{\mathcal C}}
\newcommand{\wt}{{\rm wt}}
\newcommand{\mO}{{\mathcal O}}

\begin{document}

\title{Quantum Codes from High-Dimensional Manifolds}

\author{M.~B.~Hastings}
\affiliation{Station Q, Microsoft Research, Santa Barbara, CA 93106-6105, USA}
\affiliation{Quantum Architectures and Computation Group, Microsoft Research, Redmond, WA 98052, USA}

\begin{abstract}
We construct toric codes on various high-dimensional manifolds.  Assuming a conjecture in geometry we find families of 
quantum CSS stabilizer codes on $N$ qubits with logarithmic weight stabilizers and distance $N^{1-\epsilon}$ for any $\epsilon>0$.
The conjecture is that there is a constant $C>0$ such that for any $n$-dimensional torus ${\mathbb T}^n={\mathbb R}^n/\Lambda$, where $\Lambda$ is a lattice, the least volume unoriented $n/2$-dimensional surface (using the Euclidean metric) representing nontrivial homology has volume at least $C^n$ times the volume of the least volume $n/2$-dimensional hyperplane representing nontrivial homology; in fact, it would suffice to have this result for $\Lambda$ an integral lattice with the surface restricted to faces of a cubulation by unit hypercubes.
The main technical result is an estimate of Rankin invariants\cite{rankin} for certain random lattices, showing that in a certain sense they are optimal.
Additionally, we construct codes with square-root distance, logarithmic weight stabilizers, and inverse polylogarithmic soundness factor (considered as quantum locally testable codes\cite{qltc}).
We also provide an short, alternative proof that the shortest vector in the exterior power of a lattice may be non-split\cite{coulangeon}.
\end{abstract}
\maketitle

\section{Introduction}
Quantum CSS stabilizer codes\cite{css} can be understood in terms of homology\cite{csshomology1,csshomology2,csshomology3}, and different manifolds provide a rich source of
different codes. The two-dimensional toric code\cite{csshomology1,csshomology2} and four-dimensional toric code\cite{4dtoric} are commonly considered examples; they are code families based on families of cellulations of a two and four dimensional tori.
Other manifolds\cite{fml} provide other interesting properties, such as greater distance, discussed below.
In this paper, we consider families of codes based on high dimensional manifolds.

We begin by considering some parameters that quantify a CSS code.
The elementary degrees of freedom of a CSS codes are qubits (or, more generally, qudits, for some $d\geq 2$).  Let there be $N$ such qudits so that the Hilbert space has dimension $d^N$.
CSS codes can be parametrized by several parameters, which we write as $[[N,K,D,W]]$.  Here $N$ is the number of qudits.  $K$ is the number of encoded
qudits, so that the code has a code space which is a subspace of dimension $d^K$.
$D$ is the ``distance" of the code, defined below, while $W$ is the ``weight" of the stabilizers, defined also below.
Generally speaking, larger $K$ and $D$ is desirable, while smaller $W$ is also desirable (this discussion of desirability of certain values of the parameters
ignores other questions like the ability to efficiently decode or encode states, which is a completely separate discussion that we do not consider in this paper).

The best families of quantum codes obtained thus far have significantly worse scaling than the corresponding scaling for classical linear codes.
Families of classical codes exist with $K=\Theta(N), D=\Theta(N), W=\mO(1)$ (so-called low density parity check codes provide such an example\cite{ldpc}).  If we set $W=\mO(1)$, then the largest known distance for a quantum code family is $\Theta(\sqrt{N \log(N)})$as in Ref.~\onlinecite{fml}, while
if we want $D=\Theta(N)$, then the lowest known weight is $W=\Theta(\sqrt{N})$ as in Ref.~\onlinecite{bh}.
These parameters refer to stabilizer codes; if one allows subsystem codes\cite{subsys}, then it is possible to achieve $D=\Theta(N^{1-\epsilon}),W=\mO(1)$ for $\epsilon=\mO(1/\sqrt{\log(N})$ as in Ref.~\onlinecite{subsyslin}, but now the parameter $W$ does not refer to the weight of a set of commuting stabilizers but rather the to weight of a set
of generators of the ``gauge group" and these generators need not commute with each other.  If one requires that the stabilizer group be generated by local commuting operators, then currently no advantage is known for a subsystem code.
Another notable stabilizer code family achieves $k=\Theta(N),d=\Theta(\sqrt{N}),w=\mO(1)$ and has efficient an efficient local decoding algorithm\cite{tz}.

In this paper, we construct code families that, assuming a conjecture in geometry, have almost linear distance and logarithmic weight generators.  We review various concepts before giving an overview of the paper.

\subsection{Review of CSS Codes and Relation to Homology}
The code
subspace is the subspace of the $d^N$-dimensional Hilbert space which is in the $+1$ eigenspace of several ``stabilizers".  These stabilizers are of two types, called
``X-type" and ``Z-type".  The $Z$ operator on the $d$-dimensional Hilbert space of a single qudit is the operator
\be
Z=\begin{pmatrix} 1 & \\ &\exp(\frac{2\pi i}{d})\\ && \exp(\frac{4\pi i}{d}) \\ &&&\ldots\end{pmatrix},
\ee
while the $X$ operator is the operator
\be
X=\begin{pmatrix} 0 & 1 \\ & 0 & 1 \\ && 0 & 1 \\ &&&\ldots \\ 1 & 0 \ldots \end{pmatrix}.
\ee
We write $Z_i$ or $X_i$ to indicate the operator $Z$ or $X$ acting on qudit $i$, tensored with the identity on all other qudits.
Then, a Z-type stabilizer is the tensor product of $Z$ operators on some qubits, possibly raised to integer powers.  Such a Z-type stabilizer might be written, for
example, $Z_1 Z_3^2$ to indicate that it is the tensor product of $Z$ on qubit $1$ with the square of $Z$ on qubit $3$.
These exponents all can be taken in the range $1,2,...,d-1$; if an operator on a given qubit is raised to power $0$, we simply do not write it when
writing the $Z$ stabilzer.
The X-type stabilizers are similar, with $Z$ replaced by $X$.

We encode the Z-type stabilizers in a matrix that we denote $\partial_2$.  This matrix has $N$ rows and has one column per Z-type stabilizer.
The entries of the matrix are over the field $\mF_d$.  The entry in the $i$-th row and $j$-th column indicates which power of $Z_i$ appears in the $j$-th stabilizer; thus, for example, for the stabilizer $Z_1 Z_3^2$, the first row in the corresponding column would have a $1$ and the third row would have a $2$ and all other rows would be zero.
We encode the X-type stabilizers also in a matrix, denoted by $\partial_1$.  This matrix has $N$ columns and one row per X-type stabilizer, again
with the entries over the field $\mF_d$.   The entry in the $i$-th row and $j$-th column indicates which power of $X_j$ appears in the $i$-th stabilizer

A final requirement on CSS codes is that the stabilizers commute with each other.  Any pair of Z-type stabilizers trivially commute, as do any pair of
X-type stabilizers.  The requirement that the Z-type stabilizer commute with the X-type stabilizers can be simply expressed in terms of $\partial_2,\partial_1$ as
\be
\partial_1 \partial_2=0.
\ee
This requirement is equivalent to saying that there is a chain complex
$$\mC_2 \stackrel{\partial_2}{\rightarrow} \mC_1 \stackrel{\partial_1}{\rightarrow} \mC_0,$$
where $\mC_2,\mC_1,\mC_0$ are vector spaces over $\mF_d$, with basis elements in one-to-one correspondence with Z-type stabilizers, qudits, and X-type stabilizers, respectively.  We have ${\rm dim}(C_1)=N$.

The number of encoded qudits $K$ is given by the first Betti number, which is equal to $N-{\rm dim}(\mC_2)-{\rm dim}(\mC_0$ assuming that all stabilizers
are independent of each other (i.e., that the columns of $\partial_2$ are linearly independent, as are the rows of $\partial_1$).

The distance $D$ is defined as follows.  Let us say that an operator $O$
is a Z-type logical operator if it is a tensor product of $Z$ operators on qudits which commutes with all X-type stabilizers and which is not itself
a product of Z-type stabilizers.  In the language of homology, such an operator is a representative of a nontrivial first homology class; 
write $$O=\prod_i Z_i^{a_i},$$ where the product ranges over all qudits and $a_i$ are in $\mF_d$.
Define an $N$-component vector $v$ with entries $a_i$, so that the requirement that $O$ commutes with all X-type stabilizers is that $\partial_1 v=0$,
while the requirement that $O$ not be a product of Z-type stabilizers is that $v$ is not in the image of $\partial_2$.
An X-type logical operator is defined similarly, with $Z$ and $X$ integerchanged everywhere in the definition.
The weight of a Z-type (or X-type) logical operator $O$ is defined to be the number of qudits $i$ such that $Z_i$ (or $X_i$)appears in $O$ raised to a nonvanishing power mod $d$; we say that that $Z_i$ or $X_i$ is in the support of the logical operator.  We define $D_Z$ to be the minimum weight of a Z-type logical operator and $D_X$ to be the minimum weight of an X-type logical operator
and define
\be
D={\rm min}(D_X,D_Z).
\ee

We define the weight $W$ of a code to be the least integer $W$ such that every row and every column of $\partial_2$ has at most $W$ nonvanishing entries and also every row and every column of $\partial_1$ has at most $W$ nonvanishing entries.  Note that this means that not only does every stabilizer act on at most $W$ different qudits, also every qudit is acted on by at most $W$ different $Z$-type stabilizers and $W$ different $X$-type stabilizers.

We define the weight of an operator which is a product of $Z$ and $X$ operators to be the number of qudits on which the operator acts nontrivially; for example, the operator $X_1 X_3$ has weight $2$.  Thus, every stabilizer has weight at most $W$.

A vector $v$ in a vector space $\mC_k$ is called a $k$-chain (or simply, a ``chain").  If $\partial_k v=0$, then $v$ is called a $k$-cycle.
The weight of a vector is defined to be the number of nonzero entries in the vector.

\subsection{CSS Codes from Manifolds and Systolic Freedom}
Conversely, just as one can define a chain complex from a CSS code, one can use a chain complex to define a CSS code.  Given any chain complex
over some field $\mF_d$, one can define a qudit CSS code: choose any vector space in the chain complex to correspond to the qudits, and then the vector spaces of one higher and one lower dimension correspond to the Z-type and X-type stabilizers.
For example, given a triangulation (or cubulation or other discretization) of a four dimensional manifold one can define a chain complex
$$\mC_4 \stackrel{\partial_4}{\rightarrow} \mC_3 \stackrel{\partial_3}{\rightarrow} \mC_2 \stackrel{\partial_2}{\rightarrow} \mC_1 \stackrel{\partial_1}{\rightarrow} \mC_0,$$
where the basis elements of $\mC_k$ correspond to $k$-cells.  Then, one can choose any integer $q$ and let the qudits correspond to the
$q$-cells and the Z-type stabilizers correspond to $(q+1)$-cells and the X-type stabilizers correspond to $(q-1)$-cells.
The case $q=2$ is the familiar four-dimensional toric code of Ref.~\onlinecite{4dtoric}, while the cases $q=0,4$ are classical repetition codes (Ising models) in the $Z$ or $X$ basis, respectively.

Defining CSS codes from manifolds has several nice advantages.  For one, often the distance of the code can be translated into geometric properties of the manifold and (up to some technical details that we discuss below) it can be geometrically interpreted as the least possible volume of a $q$-dimensional
surface in a nontrivial homology class. Similarly, if the triangulation has a bounded local geometry, then this gives a bound on $W$.

Naively, it might seem that such constructions will not be able to obtain a better-than-square-root distance, i.e. $D=\Omega(\sqrt{N})$.
We now give some intution for this naive belief, and give a more detailed discussion of the relation between volume and number of qudits in one particular example, as it will be
useful later.
Consider an $n$-dimensional torus constructed from a hypercube of length $\ell$ on each side for some integer $\ell$ by gluing the opposite
faces together.  Introduce coordinates $(x_1,\ldots,x_n)$.  Discretize the torus by hypercubes of unit length in the obvious way, so that the $0$-cells are at integer values of the coordinates.
In this case, the volume of the torus  is equal the number of hypercubes in the discretization, which equals $\ell^d$.  The number of qudits is given by
$$N=\ell^d {d \choose q},$$
while
$$D_Z=\ell^q, \quad D_X=\ell^{d-q}.$$
To see that $D_Z\leq \ell^q$, one can pick any $q$-dimensional plane where $q$ of the coordinates assume arbitrary values and the other coordinates are held fixed at integer values; then, the product of $Z$ over the $q$-cells in this plane give a logical operator.  We omit the proof that this upper bound for $D$ is tight in this case.  The value of $D_X$ is given by picking any $(d-q)$-dimensional plane on the dual lattice and then taking the product of $X$ over the the $q$-cells that intersects this plane also gives a logical operator.

Choosing the optimal value, $q=d/2$ still leads only to $D=\Theta(\sqrt{N})$.
Varying the geometry of the torus by changing the aspect ratio (i.e., keeping the sides of the torus orthogonal to each other but changing the relative lengths) does not lead to any improvement.

However, this naive belief is false.  ``Systolic freedom" is the term for a concept due to Gromov\cite{sysfree}, that one may have manifolds for which the product of the $q$-systole (the least volume surface representing a nontrivial element of $q$-th homology) times the $(d-q)$-systole may be arbitrarily larger than the volume of the manifold.  This phenomenon was originally observed for integer homology (corresponding to qudit quantum codes with large $d$), while only later in Ref.~\onlinecite{fml} was it constructed for $\mZ_2$ homology.

\subsection{Overview of Paper}
In the original construction of systolic freedom\cite{sysfree}, the topology of the manifold was held fixed and the metric was varied to obtain a diverging ratio, while in the $\mZ_2$ case\cite{fml}, the topology of the manifold was varied to obtain a diverging ratio.  In this paper, we consider instead a family of manifolds with different dimension.  Most of the paper is devoted to considering tori ${\mathbb R}^n/\Lambda$ for certain random lattices $\Lambda$.  In section \ref{defns} we make various definitions of the random lattices and define Rankin invariants.  In section \ref{overview} we give an overview of the construction and present a geometric conjecture \ref{conj1} and state theorem \ref{mainth} that, assuming the conjecture, there exist quantum CSS codes with logarithmic weight and almost linear distance.  In section \ref{rankininvar} we prove lower bounds on the Rankin invariant of certain random lattices, which is the main step in proving theorem \ref{mainth}.  In section \ref{calibrationsec} we discuss some obstacles to proving even a weaker form of conjecture \ref{conj1} (involving oriented surfaces) and we consider shortest vectors in the exterior product of a lattice.  Finally, in section \ref{qltcsec} we give some alternative constructions which have only square-root distance but which have inverse polylogarithmic soundness parameters as quantum locally testable codes\cite{qltc}.

To give some motivation to our lattice construction, consider the two-dimensional toric code.  On a square lattice with length $\ell$ on each side, there are $2\ell^2$ qubits and the distance $\ell$.
Suppose we ignore the details of the cellulation and take an arbitrary torus ${\mathbb R}^2/\Lambda$, pretending that the number of qubits is equal to the area ($\ell^2$) and the distance is equal to the shortest vector in the lattice $\Lambda$.
Then, a slightly better geometry than the square lattice would be to take the hexagonal lattice, as the ratio of the square of the length of the shortest vector to the area of the torus is equal to $2/\sqrt{3}$ rather than $1$.  This is only a slight constant improvement over the square lattice.  However, in higher dimensions, the shortest vector in lattice $\Lambda$ can be roughly $\sqrt{n}$ longer than the $1/n$ power of the volume of the torus ${\mathbb R}^n/\Lambda$.  Further, if we consider least volume surfaces representing nontrivial homology for $q>1$, then larger improvements are possible (at least for surfaces which are hyperplanes).  This motivates our construction and the consideration of so-called ``Rankin invariants"\cite{rankin}.

\section{Random Lattices and Definitions}
\label{defns}
Consider a so-called LDA lattice\cite{lda,lda2} as follows.  We pick a prime $p$.  We will construct a lattice which is a subset of $\mZ^n$ for some even $n$.
We first construct a linear code over field $\mF_p^n$.  We define this code by a ``code generator matrix" $G$ which is an $n$-by-$k$ matrix such that the {\it column} vectors are a basis for the codewords. (We explicitly call it a ``code generator matrix" rather than just a ``generator matrix", as we will also consider lattice generator matrices later.)
Usually in coding theory, it is instead conventional to let the {\it rows} of a code generator matrix be the basis for a code, but to maintain consistency with notation we use later, we instead
use the {\it columns} as the basis.
 We choose
the entries of $G$ independently and uniformly from $\mF_p$.  

We will be interested in taking $n$ large at fixed ratio $k/n<1$.
With high probability  (i.e., with probability tending to $1$ as $n\rightarrow \infty$ with $k/n$ fixed), $G$ is non-degenerate (see next paragraph).  Assuming that $G$ is indeed non-degenerate, one can find a permutation of the rows such that $G$ is in the form
$$G=\begin{pmatrix} A \\ B \end{pmatrix}$$ where $A,B$ are
$k$-by-$k$ and $(n-k)$-by-$k$ matrices with $A$ non-degenerate.  Then, since $A$ is non-degenerate 
there exists a sequence of elementary column operations that brings $A$ to the identity matrix, where for
a matrix over $\mF_p^n$ an elementary column operation is one of: adding one column to another, multiplying a column by any nonzero element of the field, or interchanging two columns.
These column operations bring
$G$ to the form
$$G=\begin{pmatrix} I \\ C \end{pmatrix},$$ where $I$ is the $k$-by-$k$ identity matrix and $C$ is some $(n-k)$-by-$k$ matrix.  Since the entries of $C$ are
obtained by applying these column operations to the entries of $B$, the entries of $C$
are chosen independently of each other and uniformly from $\mF_p$, i.e., applying any elementary column operation to an ensemble of matrices with entries chosen uniformly and
independently leaves this ensemble invariant.
This is the form of $G$ that we work with in the rest of this section.
\begin{definition}
Let the lattice $L_0$ be the set of points $x_1,...,x_n$ in $\mZ^n$ such that the vector $(x_1 \mod p,...,x_n \mod p)$ is in the linear code defined by $G$.
\end{definition}

We now show that with high probability, $G$ is non-degenerate. With probability $1-(1/p)^n$, the first column of $G$ has a nonzero entry.  By elementary column operations, adding a multiple of the first column to other columns, we can set all other columns equal to zero
in the first row for which the first column has a nonzero entry.  Then, with probability $1-(1/p)^{n-1}$, the second column has a nonzero entry in some other row.  Add a multiple of the second column to the third, fourth,... column to set them equal to zero in the first row for which the second column has a nonzero entry.
Continuing in this fashion, the
probability that $G$ is non-degenerate is $(1-(1/p)^n)(1-(1/p)^{n-1})\ldots(1-(1/p)^{n-k+1}$ which indeed is $1-o(1)$.

The lattice $L_0$ is the set of integer linear combinations of the columns of $G$ (interpreted as vectors of integers, rather than as vectors of elements of $\mF_p$) and of the $n$ vectors with a $p$ in one coordinate and zeroes elsewhere.
Then, the lattice $L_0$ is the set of integer linear combinations of the columns of the matrix
$$\begin{pmatrix} I & pI & 0 \\ C & 0 & pI \end{pmatrix},$$ where the row blocks have sizes $k$ and $n-k$ respectively, while the column blocks have
sizes $k$, $k$, $n-k$, and respectively, and where $I$ is the identity matrix of appropriate size.
However, any integer linear combination of column vectors of $\begin{pmatrix} pI \\ 0 \end{pmatrix}$ is also an integer linear combination of column vectors of 
$$\begin{pmatrix} I & 0 \\ C & pI \end{pmatrix}.$$ To see this, consider any vector of integers $\vec y=(y_1,...,y_k)$.  Then,
\begin{eqnarray}
\begin{pmatrix} pI \\ 0 \end{pmatrix} \vec y&=&\begin{pmatrix} p \vec y \\ \vec 0\end{pmatrix}\\ \nonumber &=&\begin{pmatrix} I \\ C \end{pmatrix} \begin{pmatrix}p \vec y \\ \vec 0\end{pmatrix}-\begin{pmatrix} 0 \\ pI \end{pmatrix}
\begin{pmatrix} 0 \\ C\vec y \end{pmatrix}.
\end{eqnarray}
Thus, $L_0$ is the set of integer linear combinations of columns of the matrix
$$B_0=\begin{pmatrix} I & 0 \\ C &pI \end{pmatrix}.$$

A matrix $B$ such that the lattice is the set of integer combinations of columns of $B$
is called a generating matrix for the lattice.
Two different generating matrices $B_1,B_2$ define the same lattice if and only if $B_1=B_2 T$ where $T$ is an integer matrix such that $T^{-1}$ also is an integer matrix.  In this case, the matrix $B_1$ can be turned into the matrix $B_2$ by a sequence of elementary column operations where an elementary column operations is one of: adding one column to another, changing the signs of all entries in a column, or interchanging two columns.

Given a lattice $L$ with generating matrix $B$ which is an $n$-by-$k$ matrix, such that $B$ has rank $k$, we define the volume of the lattice
to equal $\vol(L)={\rm det}(B^\dagger B)^{1/2}$.
If $k=n$, then $\vol(L)=|{\rm det}(B)|$.
\begin{definition}
Given any linearly independent set of vectors $x_1,...,x_k$ in $\mZ^n$ (or more generally in ${\mathbb R}^n$) we define their volume $\vol(x_1,...,x_k)$ to be the volume
of the lattice generated by the $n$-by-$k$ matrix with columns $x_1,...,x_k$.
\end{definition}

This matrix $B_0$ is upper triangular and so ${\rm det}(B_0)$ is easily computed:
\be
\vol(L_0)=|{\rm det}(B_0)|=p^{n-k}.
\ee

\begin{definition}
An ``integral lattice" is defined to be a lattice whose generating matrix has integer entries.  A ``primitive lattice" is defined to be an integral lattice such that there is no other
integral lattice of the same rank properly containing it.  Equivalently, there is no integral lattice which spans the same subspace and properly contains it.
\end{definition}
Example: in two dimensions, the lattice generated by the vector $(2,1)$ is primitive, while that generated by $(4,2)$ is not.

Unless specified, all lattices will be in $n$ dimensions.
We use $|\ldots |$ to denote the $\ell_2$ norm of a vector.

Finally, we define the Rankin invariant.
\begin{definition}
The Rankin invariant $\gamma_{n,m}(L)$ for a lattice $L$ with rank $n$  is defined to be
\be
\gamma_{n,m}(L)={\rm min}_{\stackrel{v_1,...,v_m \in L}{\vol(v_1,...,v_m) \neq 0}} \Bigl( \frac{\vol(v_1,...,v_m)}{\vol(L)^{m/n}} \Bigr)^2.
\ee
The square in the above definition is included for historical reasons.  The factor $m/n$ in the exponent of $L_0$ is such that the
invariant is unchanged under rescaling the lattice $L$ by any constant factor.  In the case $m=1$, the Rankin invariants $\gamma_{n,1}(L)$ is related to the length of the shortest vector:
$\gamma_{n,1}(L)={\rm min}_{x \in L, x \neq 0} \frac{|x|^2}{\vol(L_0)^{2/n}}$.
Clearly, $\gamma_{n,n}(L)=1$ for all $L$.
\end{definition}

The Rankin invariant $\gamma_{n,1}(L)$ is related to the length of the shortest vector in the lattice.  To understand the higher Rankin invariants, consider a set of vectors $v_1,...,v_m \in L$ with $\vol(v_1,...,v_m)\neq 0$.  Consider the torus ${\mathbb R}^n/L$.
The $m$-dimensional hyperplane spanned by $v_1,...,v_m$ represents a nontrivial integer homology class and has an $m$-dimensional volume (using the Euclidean metric) equal to $\vol(v_1,...,v_m)$.

\section{Overview of Construction: Conjectures and Main Result on Distance}
\label{overview}
We
will consider a family of CSS codes obtained by choosing a fixed $p>1$ and taking LDA lattices with $k=n/2$ from the random ensemble above, for all (even) values of $n$.
With high probability, this lattice has rank $n$.
Given the lattice, we take a cellulation of the lattice by hypercubes of length $1$ on each side.
Then, we consider a qubit toric code on this cellulation with degrees of freedom on $q$-cells for $q=n/2$.
Then, the number of $q$-cells is equal to
\be
N={n \choose n/2} p^{n/2}.
\ee

The distance of the code is equal to the weight of the least weight logical $X$ or $Z$ operator.
The vector corresponding to such an operator represents nontrivial homology or cohomology with ${\mathbb Z}_2$ coefficients.
We conjecture that:
\begin{conjecture}
\label{conj1}
There exists a constant $C>0$, such that for any $n$-dimensional integer lattice $L$, for the toric code obtained by the cellulation using integer
hypercubes and degrees of freedom on $q$-cells for $q=n/2$,
the distance is lower bounded by $C^n {\rm min}_{\stackrel{v_1,...,v_q \in L}{\vol(v_1,...,v_q) \neq 0}} \vol(v_1,...,v_q)=C^n \vol(L)^{q/n} \gamma_{n,q}(L)^{1/2}$.
\end{conjecture}

Let us motivate this conjecture.  The least volume hyperplane representing nontrivial homology has volume equal to the Rankin invariant.
This hyperplane need not lie on the $q$-cells that we have chosen.  We can deform the hyperplane  to get a surface that lies on the $q$-cells using the Federer-Fleming deformation theorem\cite{ff}: this theorem is based on deforming the surface to lie on the $(n-1)$-skeleton (i.e., the $(n-1)$-dimensional faces of the hypercubes of unit size), then on the $(n-2)$-skeleton, and so on, iteratively, until the surface likes on the $q$-skeleton.  The deformation to move surface from the $m$-skeleton to the $(m-1)$-skeleton is done by choosing a point randomly in an $m$-dimensional hypercube and then projecting the surface outwards from that point to the boundary.
This deformation may increase the volume, but that is fine: what we are considered with is lower bounding the volume.

However, it is not clear the the optimal operator is obtained by such a deformation procedure starting from a hyperplane.  There may be, for example, unoriented
chains which are not hyperplanes but which represent nontrivial homology and have much smaller volume than the least volume hyperplane.
The conjecture is that such surfaces can have at most exponentially smaller (i.e., smaller by a factor $C^n$) volume.

Conjecture \ref{conj1} considers the distance of the code, which is equal to the least volume of a $\mZ_2$
cycle representing nontrivial homology.  The cycles are in the chain complex obtained from the cellulation using hypercubes.
One may be tempted to make a (possibly stronger) conjecture that a similar inequality holds for more general chains, such as polyhedral chains.
In this regard, we remark that the
possible increase in volume from the Federer-Fleming deformation theorem may be superexponentially large: the upper bound is at most $2n^{n/2} {n \choose n/2}$ (see Ref.~\onlinecite{ffencyc}).

We prove that:
\begin{theorem}
\label{mainth}
Assume that conjecture \ref{conj1} holds.  Then, for any $\epsilon>0$, there exists a family of quantum CSS codes
on $N$ qubits with distance $D=\Omega(N^{1-\epsilon})$ and weight $w=\mO(\log(N))$ and
with $\Theta(N^\delta)$ encoded qubits, where $\delta>0$ ($\delta$ depends on $\epsilon$).
\end{theorem}

This theorem will follow from a corollary of theorem \ref{countingthm}, which implies that for any constant $c<1/\sqrt{2\pi e}$, with high probability we have
${\rm min}_{\stackrel{v_1,...,v_q \in L}{\vol(v_1,...,v_q) \neq 0}} \vol(v_1,...,v_q) \geq (cp)^{n/2}$.
Hence, with high probability, $d \geq (cC^2 p)^{n/2}$.  Since $N={n \choose n/2} p^{n/2} \leq (4p)^{n/2}$,
with high probability we have
$$d \geq (cC^2 p)^{\log_{4p}(N)}=N^{\frac{\log(cC^2p)}{\log(4p)}}.$$
Fixing $c$ to be any constant slightly smaller than $1/\sqrt{2\pi e}$, we find that for any $\epsilon>0$ that for all sufficiently large $p$ we have
$$1-\epsilon \leq \frac{\log(cC^2p)}{\log(4p)}$$
so that $d\geq N^{1-\epsilon}$.

We have $w=\mO(d)=\mO(\log(N))$.

The number of encoded qubits is equal to ${n \choose k}=2^{(1-o(1)) n}=2^{2(1-(o(1))\log_{4p}(N)}=N^{2(1-o(1))/\log(4p)} \equiv N^{\delta}$.

The main work will be theorem \ref{countingthm}, to lower bound the Rankin invariant for this class of lattices.
The reader may wonder why we introduce this class of lattices, instead of re-using previous results which show that
there exist random lattices with a large Rankin invariant, $\gamma_{n,n/2}(L) \geq (\frac{k}{12})^{n/4}$.  See theorem 3 in Ref.~\onlinecite{blockwise}.
The reason is that the random lattices constructed there need not be integral lattices and so we do not have such an obvious
cell decomposition to place on the lattices.
We comment later on the relationship between the Rankin invariant for our lattice (which depends on $n,p$) and the invariant
in Ref.~\onlinecite{blockwise}; this requires considering how large $n$ needs to be compared to $p$ in our construction.

Note that we choose $p$ large so that the exponentially growing factor, $\approx 2^n$, arising from the factor ${n \choose n/2}$ in the number of cells will be polynomially smaller than the volume $p^{n/2}$.  We have
$2^n=(p^{n})^{1/\log_2(p)}$.

We remark that similar code constructions can be made by choosing degrees of freedom on $q$-cells for $q\neq n/2$, taking $n$ large at a fixed ratio $q/n$.  In this case, a natural generalization of conjecture \ref{conj1} is to assume that 
there is a constant $C$ such that $d_Z \geq 
C^n \vol(L)^{q/n} \gamma_{n,q}(L)^{1/2}$ and $d_X \geq
C^n \vol(L)^{(n-q)/n} \gamma_{n,n-q}(L)^{1/2}$.
Assuming this conjecture, our construction would give a code with $d_X d_Z$ polynomially larger than $N$.

\section{Rankin Invariants}
\label{rankininvar}
In this section, we will prove lower bounds on the Rankin invariants\cite{rankin} $\gamma_{n,m}(L_0)$ of $L_0$.  
The proof uses the probabilistic method; in particular, we use the first moment method.  To motivate the proof, let us first sketch a proof method for $\gamma_{n,1}(L_0)$; then, we give a sketch a possible extension of the proof method to $\gamma_{n,m}(L_0)$ and explain some difficulties with this extension; finally, we outline the approach we use which is a modification of that.  First, suppose we just want to lower bound $\gamma_{n,1}(L_0)$; i.e., we wish to lower bound the shortest vector in the lattice.  This can be done by a first moment method: estimate the number of integer vectors with length less than some given length $\ell$; then, compute the probability that any given integer vector is in the lattice (this probability is $p^{-(n-k)}$ for a randomly chosen code assuming $G$ is non-degenerate); so, for sufficiently small $\ell$,  the average number of integer vectors with length less than $\ell$ in the lattice is small so it is unlikely that any integer vectors with length less than $\ell$ will be in the lattice.  One might attempt to do something similar for the Rankin invariants:
estimate the number of rank $m$ integral lattices in $n$ dimensions with volume at most $V$ and then compute the probability that
an integral lattice is in a randomly chosen linear code.  Call this rank-$m$ lattice $K$ and call its generating matrix $M_K$.
In fact, Ref.~\onlinecite{schmidt0} provides
asymptotic estimates (large $V$) for the number of such lattices $K$, so it might seem that one could directly use the results there
in a first moment method.
Indeed, this approach might work, but since the results of Ref.~\onlinecite{schmidt0} hold in the asymptotic limit (large $V$), some additional estimates would be needed  (we do use many results in Ref.~\onlinecite{schmidt0}).
However, the results we need are in some ways simpler than that of Ref.~\onlinecite{schmidt0} because we do not care about an exact estimate of the number of such lattices, only an upper bound.
Further, rather than applying the first moment method by estimating the number of lattices $K$ with some given volume and estimating the probability that such a lattice is in the code and then showing that the product is small for small $V$, we will apply the first moment method to each column of the generating matrix $M_K$ {\it separately} (with $M_K$ written in Hermite normal form).  That is, we first estimate (this step is exactly analogous to the discussion at the start of this paragraph regarding how to lower bound $\gamma_{n,1}(L_0)$) the probability that there is a choice for the first column which has small length and which is in the code.
Then, we estimate the probability that there is a choice for the second column which is also in the code such that the ratio of the volume of the lattice generated by the first two columns of $M_K$ to the volume of the lattice generated by the first column of $M_K$ is small.  To do this calculation, we need the concept of ``factor lattice"\cite{schmidt0}, reviewed below.  We continue in this fashion over the other columns, showing that the ratio of the volume of the lattice generated by the first $a$ rows of $M_K$ to the volume of the lattice generated by the first $a-1$ rows of $a$ is likely to be large, for each $a=2,3,\ldots$.

\subsection{Counting Points}
Let $V_d(r)$ denote the volume of a ball of radius $r$ in $d$ dimensions:
\be
V_d(r)=\frac{\pi^{d/2}}{\Gamma(\frac{d}{2}+1)}r^d.
\ee

Given a rank-$l$ lattice $L$ which spans some space $E$, we define the Voronoi cell to be the set of points $y$ in $E$ such that $|y| \leq |y-v|$ for all lattice points $v \neq 0$.
The $l$-dimensional volume of the Voronoi cell is equal to $\vol(L)$.

\begin{definition}
Given a lattice $L$, let $N(L,z,r)$ denote the number of points in lattice $L$ within distance $r$ of some given point $z$.
\end{definition}
\begin{lemma}
\label{numpoints}
Let $L$ be a rank-$l$ lattice in $d$ dimensions which spans some space $E$.  Suppose the diameter of the Voronoi cell of $L$
 is bounded by some given $D$.  Then, for any $z,r$, 
\be
N(L,z,r) \leq \frac{1}{\vol(L)} V_l(r+D).
\ee
\begin{proof}
For every point $x\in L$ within distance $r$ of $z$, let $T_x$ be the set of points $y\in E$ such that $y-x$ is in the interior of the Voronoi cell.
The sets $T_x$ are non-overlapping and each has $l$-dimensional volume $\vol(L)$.  So, the volume of $\cup_{x,|x-z|\leq r} T_x$ is equal to $N(L,z,r) \vol(L)$.
Every $x$ is within distance $r$ of $z$ and so every point in $\cup_{x,|x-z|\leq r} T_x$ is within distance $r+D$ of $z$, so $N(L,z,r) \vol(L) \leq V_k(r+D)$.
\end{proof}
\end{lemma}

We make some more definitions.
\begin{definition}
Given a rank-$l$ lattice $L$ spanning a subspace $E$, the polar lattice $L^P$ is the lattice of all vectors in $E$ which have integral inner products with all vectors
in $L$.  The polar lattice also has rank $l$ and $\vol(L^P) \vol(L)=1$.
\end{definition}

\begin{definition}
Let $\Gamma_0^n$ be the rank-$n$ lattice in $n$ dimensions consisting of all vectors for which all coordinates are integral.  
\end{definition}

\begin{definition}
Given a primitive lattice $L$ spanning subspace $E$, the orthogonal lattice $L^\perp$ consists of all vectors in $\Gamma_0^n$ with vanishing inner product
with all vectors in $L$.
\end{definition}

\begin{definition}
Let $L$ be a rank-$l$ primitive sublattice of $\Gamma_0^n$ and let $E$ be the subspace spanned by $L$.  Let $\pi$ project onto the orthogonal complement of $E$, which we write $E^\perp$.
Let $\pi(\Gamma_0^n) \equiv \Gamma_0^n/L$.  Then, $\Gamma_0^n/L$ is also a lattice, called the factor lattice.  It has rank $n-l$\end{definition}

We have\cite{schmidt0}
\be
\label{volinvert}
\vol(L) \vol(\Gamma_0^n/L)=1.
\ee
This equation follows from this lemma:
\begin{lemma}
\be
\Gamma_0^n/L=((L)^\perp)^P.
\ee
\begin{proof}
See Ref.~\onlinecite{schmidt0}.
\end{proof}
\end{lemma}

\begin{lemma}
\label{diamboundlemma}
Let $L$ be a rank-$l$ primitive sublattice of $\Gamma_0^n$.  Let $\pi$ and $\Gamma_0^n/L$ be as above.  Then, the diameter of the Voronoi cell of $\Gamma_0^n/L$ is bounded by $\sqrt{n-l}$.
\begin{proof}
Since $L$ has rank $l<n$, there must be some vector $w_1$ which has a $1$ in one coordinate and zeroes in all other coordinates (i.e., $w_1$ is of the form $(0,\ldots,0,1,0,\ldots,0)$) which is not in $E$.  Then, since the span of $E$ and $w_1$ has dimension $l+1$, if $k<n-1$, there must be some other vector $w_2$ of the same form which is not in the span of $E$ and $l_1$.  Proceeding in this fashion, we construct vectors $w_1,...,w_{n-l}$, all of which have zeroes in all but one coordinate and a $1$ in that coordinate.  The vectors $\pi(w_i)$
span $E^\perp$.  So, every point $y$ in $E^\perp$ can be written as a linear combination $y=\pi(\sum_i a_i w_i)$.  If the $a_i$ are integer, then $y$ is a lattice point in $\pi(\Gamma_0^n)$.
Every linear combination $\sum_i a_i w_i$ is within distance $(1/2)\sqrt{n-l}$ of some linear combination $\sum_i b_i w_i$ with integer $b_i$ (to see this, simply round all $a_i$ to the nearest integer).
Since the norm does not increase under projection, every $\pi(\sum_i a_i w_i)$ is also within distance $(1/2) \sqrt{n-l}$ of some $\pi(\sum_i b_i w_i)$ for integer $b_i$ and
hence every point in $E$ is within distance $(1/2) \sqrt{n-l}$ of a lattice point.
\end{proof}
\end{lemma}
We remark that
the lattice with basis vectors $\pi(w_i)$ may not include all points in $\pi(\Gamma_0^n)$; as an example, consider $l=1$ and $n=2$ and let $L$ be the lattice with basis vector $(2,1)$ and let $w_1=(0,1)$.  The vector $\pi((1,0))$ is then not included in the lattice with basis vector $\pi(w_1)$.

\begin{lemma}
\label{pointsboundlemma}
Let $L$ be a rank-$l$ primitive sublattice of $\Gamma_0^n$.  Let $\pi$ and $\Gamma_0^n/L$ be as above.  The number of points in $\Gamma_0^n/L$ within
distance $r$ of the origin is bounded by
\be
N(\Gamma_0^n/L,0,r) \leq \vol(L) V_l(r+\sqrt{n-l}).
\ee
\begin{proof}
This follows from lemmas \ref{numpoints},\ref{diamboundlemma} and Eq.~(\ref{volinvert}).
\end{proof}
\end{lemma}

\subsection{Hermite Normal Form For Lattices}
Consider a rank-$m$ integral lattice $K$ in $n$ dimensions.  If this lattice has basis vectors $v_1,...,v_m$, we write an $n$-by-$m$ matrix $M_K$ whose columns are these basis vectors.  We label the rows of the matrix by integers $1,\ldots,n$ and label the columns by integers $1,\ldots,m$.
Such a matrix is called a lattice generator matrix for the lattice.
Then, the set of points in the integral lattice is the image under $M_K$ of $\Gamma_0^{k}$.
By a sequence of column operations (adding one column of $M_K$ to another column, which does not change the image, or changing the sign of a column, which also does not change the image),
we can bring always bring the matrix $M_K$ to so-called ``Hermite normal form"; further, there is a unique matrix $M_K$ in Hermite normal form which generates $K$.

Our definition of Hermite normal form differs from that of other authors because we will {\it reverse} the order of columns and {\it reverse} the order of rows
compared to the usual order.  This is because we will be doing induction later and with the reversed order of columns, the
notation will be much more natural later.  See Eq.~(\ref{example}) for an example of Hermite normal form below.

\begin{definition}
\label{HNF}
A matrix
$M_K$ is said to be in Hermite normal form if for every column $j$ there is a row $i_j$ with $1 \leq i_1 < i_2 < \ldots <i_m\leq n$ such that the entries of $M_K$ obey:
\be
\label{zeroes}
i>i_j \quad \rightarrow \quad (M_K)_{i,j}=0
\ee
and
\be
\label{pos}
(M_K)_{i_j,j}>0,
\ee
and 
\be
\label{Sformnew}
l > j \quad \rightarrow \quad
0 \leq (M_K)_{i_j,l} < (M_K)_{i_j,j}.
\ee

We say that ``the first $a$ columns of $M_K$ are in Hermite normal form" if the submatrix of $M_K$ consisting of the first $a$ columns is in Hermite normal form.  In this case, for every column $j$ with $j\leq a$ there is a row $i_j$ with $1 \leq i_{1}<i_{2} <\ldots <i_a \leq n$ such that
Eqs.~(\ref{zeroes},\ref{pos},\ref{Sformnew}) hold when restricted to the case that $j \leq a$ and $l \leq a$.
\end{definition}

We introduce some notation.  This notation defines various vector spaces and vectors in terms of the matrix $M_K$; we do not explicitly write $M_K$ in the definition, but rather
the particular choice of $M_K$ should be clear in context.
The last nonzero entry in the $j$-th column occurs in the $i_j$-th row. 
Define a sequence of lattices $K_1,K_2,...,K_k$, where $K_j$ has rank $j$ and $K_j$ is defined to be the lattice generated by the
submatrix of $M_K$ containing the first $j$ rows and the first $i_j$ columns.  Note that $K_k=K$.
Note also that if $K_a$ is primitive then $K_b$ is primitive for all $b<a$.

We
let $\vec v_{j}$ be the vector given by the first $i_j$ rows of the $j$-th column. 

This notation can be clarified with an example of $m=5,k=3$, with $i_1=2,i_2=4,i_3=5$, where we write
\be
\label{example}
M_K=\begin{pmatrix} 
(\vec v_1)_1 & (\vec v_2)_1 & (\vec v_3)_1 \\
(\vec v_1)_2 &(\vec v_2)_2 & (\vec v_3)_2 \\
0&(\vec v_2)_3 & (\vec v_3)_3 \\
0 & (\vec v_2)_4 & (\vec v_3)_4\\
0 & 0 & (\vec v_3)_5
\end{pmatrix},
\ee
with $(v_j)_i$ denoting the $i$-th entry of vector $\vec v_a$.
For this matrix to be in Hermite normal form, we have $0 \leq (\vec v_2)_2,(\vec v_3)_2 < (\vec v_1)_2$ and $0 \leq (\vec v_3)_4<(\vec v_2)_4$.

The lattice $K_j$ is a sublattice of $\Gamma_0^{i_j}$.  
We let $M_{K_j}$ be the submatrix of $M_K$ consisting of the first $j$ rows and the first $i_{j}$ columns so that $M_{K_j}$ generates $K_j$.
We
 also define a lattice $\tilde K_j$ which is
a sublattice of $\Gamma_0^{i_{j+1}}$.
The lattice $\tilde K_j$ will be the sublattice generated by the submatrix of $M_K$ consisting of the first $j$ rows and the first $i_{j+1}$ columns.
We let $M_{\tilde K_j}$ be the submatrix of $M_K$ consisting of the first $j$ rows and the first $i_{j+1}$ columns.
Hence, the last $i_{j+1}-i_j$ entries of every vector in $\tilde K_j$ are equal to $0$.

Let $\pi_j$ project onto the orthogonal complement of the span of $\tilde K_j$.

\begin{lemma}
\label{factorlemma}
Let $K$ be a rank-$m$ lattice in $n$ dimensions with generating matrix $M_K$ in Hermite normal form.  Then, there exist an $n$-by-$m$ integer matrix $M_{K^P}$ which is a lattice generating matrix in Hermite normal form (with the same $i_j$ as $M_K$)
for a primitive lattice, and an $m$-by-$m$ integer matrix $F$ which is upper triangular with positive diagonal entries such that we have
\be
\label{factoreq}
M_K=M_{K^P} F.
\ee

Further, $F,M_{K^P}$ are unique.
\begin{proof}
Let $K^P$ be a primitive lattice spanning the same space as $K$ and containing the lattice $K$.  (Note that such a primitive $K^P$ must exist and is unique: it is the lattice consisting of all integer points which are in the space spanned by $K$).
Let $K^P$ be generated by $M_{K^P}$ with $M_{K^P}$ in Hermite normal form; note that since $K^P$ is unique, $M_{K^P}$ is uniquely determined by $K$.
Then, since $K$ is contained in $K^P$, every column of $M_K$ is an integer linear combination of columns of $M_{K^P}$.
So, $M_K=M_{K^P} F$ for some integer matrix $F$.

$M_K,M_{K^P}$ must have the same $i_j$ or their columns would not span the same space.

Since $M_{K},M_{K^P}$ have the same $i_j$, it follows that $F$ is upper triangular with positive diagonal entries: restrict $M_K,M_{K^P}$ to the rows $i_1,i_2,\ldots,i_m$, giving upper triangular matrices of size $m$-by-$m$.  Call these matrices $A,B$ respectively. Then $A=BF$, so $F=B^{-1} A$.  Since $B$ is upper triangular, so is $B^{-1}$ and so is $F$.
\end{proof}
\end{lemma}

\subsection{Counting Column Choices}
\begin{definition}
A lattice $L$ is {\it consistent} with a code generator matrix $G$ if every point $(x_1,...,x_n)$ in the lattice has the property that $(x_1 \mod p,\ldots,x_n \mod p)$ is in the code defined by $G$.
A lattice generator matrix $M_L$ is consistent with a code generator matrix $G$ if the lattice generated by $M_L$ is consistent with $G$.
\end{definition}

We will use $a$ to label a column choice, $1\leq a \leq m$.
We will construct lattices $K_a$ in terms of $K_{a-1}$ and $\vec v_a$.
\begin{lemma}
\label{volumelemma}
Let $M_K$ be in Hermite normal form.  Then,
\be
\vol(K_{a})=\vol(K_{a-1}) |\pi_{a-1}(\vec v_{a})|.
\ee
\begin{proof}
Immediate from the definition of volume.
\end{proof}
\end{lemma}

Assume $K_{a-1}$ is primitve.
Then, the next lemma gives a one-to-one correspondence between vectors $\vec v_a$ obeying {\it one} of the conditions needed for Hermite normal form (the condition Eq.~(\ref{Sformnew})) and vectors in a certain factor lattice.  In lemma \ref{corresp2lemma} we consider the case that $K_{a-1}$ is not primitive.  Note that there is an additional condition on $v_a$, namely that its first entry be positive, in order for the matrix $M_{K_a}$ to be in Hermite normal form.
\begin{lemma}
\label{corresplemma}
Let the first $a-1$ columns of $M_K$ be given and 
assume that the first $a-1$ columns of $M_K$ are in Hermite normal form and
assume that $K_{a-1}$ is a primitive sublattice of $\Gamma_0^n$.
Then, there is a one-to-one correspondence between vectors $\vec v_{a}$ such that 
\be
\label{Sft}
j<a \quad \rightarrow \quad
0 \leq (M_K)_{i_j,a} < (M_K)_{i_j,j}
\ee
and
points $\vec x$ of the lattice $\Gamma_0^{i_a}/\tilde K_{a-1}$, such that
if $\vec x$ corresponds to $\vec v_a$ then $\pi_{a-1}(\vec v_a)=\vec x$.
\begin{proof}
We will show that for every $\vec x \in \Gamma_0^{i_a}/\tilde K_{a-1}$, there exists a unique $\vec v_a$ obeying Eq.~(\ref{Sft}) such that
$\pi_{a-1}(\vec v_a)=\vec x$.  This gives a map ${\cal F}$ from $\Gamma_0^{i_a}/\tilde K_{a-1}$ to vectors obeying Eq.~(\ref{Sft}).  This map is
one-to-one since distinct vectors $\vec x_1 \neq \vec x_2$ cannot both be the image of the same vector $\vec v_a$ under the map $\pi_{a-1}$.  
This map ${\cal F}$ is onto since any vector $\vec v_a$
obeying Eq.~(\ref{Sft}) is the image of $\pi_{a-1}(\vec v_a)$ under this map.

First we show existence of somer vector $\vec v_a$.  Every vector $\vec x$ in $\Gamma_0^{i_a}/\tilde K_{a-1}$ is given by $\vec x=\pi_{a-1}(\vec y)$ for some $\vec y\in \Gamma_0^{i_a}$.
For any such vector $\vec y$, we can add lattice vectors in $\tilde K_{a-1}$ so that Eq.~(\ref{Sft}) (i.e., set $\vec v_a$ equal to $\vec y$ plus some sum of lattice vectors; this can be done iteratively, so that it holds first for 
$j=a-1$, then $j=a-2$, and so on).  Adding these lattice vectors does not change the image of the result under $\pi_{a-1}$.

Now uniqueness.  Suppose that $\pi_{a-1}(\vec y)=\pi_{a-1}(\vec z)$ for $\vec y,\vec z$ being two possible choices of $\vec v_a$ such that Eq.~(\ref{Sft}) is obeyed.
Then, $\pi_{a-1}(\vec y-\vec z)=0$, so $\vec y-\vec z$ is in the span of $\tilde K_{a-1}$.  Since $K_{a-1}$ is primitive so is $\tilde K_{a-1}$ and so $\vec y-\vec z$ is in $\tilde K_{a-1}$.  
Let $M_K(i,j)$ denote the submatrix of $M_K$ containing the first $i$ rows and the first $j$ columns, so that $M_K(i_a,a-1)$ is a lattice generating matrix for $\tilde K_{a-1}$.  So, $\vec y-\vec z=M_K(i_a,a-1) \vec u$, where $\vec u\in \Gamma_0^{a-1}$.
Then, Eq.~(\ref{Sft}) requires that $\vec u=0$.  This follows inductively: if the last entry of $\vec u$ is nonzero, then it is not possible for both $\vec y$ and $\vec z$ to
obey Eq.~(\ref{Sft}) for $j=a-1$; to see this, note that then the $(a-1)$-th entries of $\vec y,\vec z$ must differ by
a positive integer multiple of $(M_K)_{i_j,j}$ and so they cannot both fall in the range $0,1,\ldots,(M_K)_{i_j,j}-1$.
So, $\vec y-\vec z$ differs by an elements of the lattice generated by $M_K(i_a,a-2)$ and so
$\vec y-\vec z=M_K(i_a,a-2) \vec u'$ for $\vec u'\in \Gamma_0^{a-2}$.  Again, the last entry of $\vec u'$ must equal zero so that Eq.~(\ref{Sft}) will be obeyed for $j=a-2$.  We continue inductively for $j=a-3,\ldots$.
\end{proof}
\end{lemma}

The next lemma is similar to the previous except that we no longer assume that $K_{a-1}$ is primitive.
\begin{lemma}
\label{corresp2lemma}
Let the first $a-1$ columns of $M_K$ be given and 
assume that the first $a-1$ columns of $M_K$ are in Hermite normal form.

Let $M_K(i,j)$ denote the submatrix of $M_K$ containing the first $i$ rows and the first $j$ columns, so that $M_{\tilde K_{a-1}}=M_K(i_a,a-1)$ is a lattice generating matrix for $\tilde K_{a-1}$.
Use lemma \ref{factorlemma}
to write
$$M_{\tilde K_{a-1}}=M_{{\tilde K^P}_{a-1}} F.$$

Then, the possible choices of $\vec v_{a}$ such that 
\be
\label{Sft2}
j<a \quad \rightarrow \quad
0 \leq (M_K)_{i_j,a} < (M_K)_{i_j,j}
\ee
 are in one-to-one correspondence with choices of tuples
$(\vec x,f_1,\ldots,f_{a-1})$, where $\vec x$ is a point
in $\Gamma_0^{i_a}/\tilde K^P_{a-1}$ and $f_1,\ldots,f_{a-1}$ are integers obeying
$0 \leq f_i < F_{i,i}$,
 such that
if $(\vec x,f_1,\ldots,f_{a-1})$ corresponds to $\vec v_a$ then $\pi_{a-1}(\vec v_a)=\vec x$.
Thus, there are ${\rm det}(F)$ distinct vectors $\vec v_a$ corresponding to $\vec x$.
\begin{proof}
We will show that for every $\vec x \in \Gamma_0^{i_a}/\tilde K_{a-1}$, there exist ${\rm det}(F)$ distinct vectors $\vec v_a$ obeying Eq.~(\ref{Sft2}) such that
$\pi_{a-1}(\vec v_a)=\vec x$.  These ${\rm det}(F)$ vectors will be labelled by $f_1,\ldots,f_{a-1}$.

First we show existence.  Every vector $\vec x$ in $\Gamma_0^{i_a}/\tilde K_{a-1}$ is given by $\vec x=\pi_{a-1}(\vec y)$ for some $\vec y\in \Gamma_0^{i_a}$.
For any such vector $\vec y$, we can add lattice vectors in $\tilde K_{a-1}$ so that Eq.~(\ref{Sft2}) will hold (this can be done iteratively, so that it holds first for 
$j=a-1$, then $j=a-2$, and so on).  Adding these lattice vectors does not change the image of the result under $\pi_{a-1}$.

Now, for each $\vec x$, let $\vec z$ be some fixed vector such that Eq.~(\ref{Sft2}) holds for $\vec v_a=\vec z$ and such that $\vec x=\pi_{a-1}(\vec z)$.
Suppose that $\pi_{a-1}(\vec y)=\pi_{a-1}(\vec z)$ for $\vec y$ some other possible choice of $\vec v_a$ such that Eq.~(\ref{Sft2}) is obeyed.
We count the number of possible choices of $\vec y$.
Then, $\pi_{a-1}(\vec y-\vec z)=0$, so $\vec y-\vec z$ is in the span of $\tilde K_{a-1}$. 
Since $\tilde K^P_{a-1}$ is primitive, $\vec y-\vec z=M_{{\tilde K^P}_{a-1}} \vec u$, where $\vec u\in \Gamma_0^{a-1}$.
There are $F_{a-1,a-1}$ possible choices for the $(a-1)$-th entry of $\vec u$.  To see this, note that $\vec y$ and $\vec z$ both
obey Eq.~(\ref{Sft2}) for $j=a-1$.  For $j=a-1$, this equation gives a constraint that the $(a-1)$-th entry of $\vec y$ must fall in the range $0,\ldots,(M_K)_{i_{a-1},a-1}-1$.  The $(a-1)$-th entry of $\vec y$ is determined by the $(a-1)$-th entry of $\vec u$ and shifting that
entry of $\vec u$ by one shifts the $(a-1)$-th entry of $\vec y$ by $(M_{\tilde K^P_{a-1}})_{i_{a-1},a-1}$.  We have
$(M_K)_{i_{a-1},a-1}=(M_{\tilde K^P_{a-1}})_{i_{a-1},a-1} F_{a-1,a-1}$ so that there are $F_{a-1,a-1}$ possible choices.
Then, given this choice of the $(a-1)$-th entry of $\vec u$, there are $F_{a-2,a-2}$ possible choices for the $(a-2)$-th entry of $\vec u$, and
so on.
\end{proof}
\end{lemma}

\begin{lemma}
\label{colcountlemma}
Let $M_K$ be a matrix in Hermite normal form which is a lattice generating matrix for a rank-$m$ integral lattice $K$ in $n$ dimensions.
Let $K_{a-1}$ be given and let $r$ be a real number.  Let $C(r,K_{a-1})$ denote the number of choices of $K_a$ such that
\be
\vol(K_{a}) \leq r K_{a-1}.
\ee
Then,
\be
C(r,K_{a-1}) \leq \vol(K_{a-1})  V_{i_a-a+1}(r+\sqrt{i_a-a+1}).
\ee

If $r<1$ then $C(r,K_{a-1})=0$.
\begin{proof}
Let $\vec v_a$ be as defined above.
By lemma \ref{volumelemma}
\be
\vol(K_a)=\vol(K_{a-1}) |\pi_{a-1}(\vec v_a)|.
\ee
so $|\pi_{a-1}(v_a)| \leq r$.
By Eq.~(\ref{pos}), the first entry of $\vec v_a$ is $\geq 1$, and since all vectors in $K_{a-1}$ vanish in the first entry, we have
$|\pi_{a-1}(\vec v_a)|\geq 1$, so indeed $C(r,K_{k-1})=0$ for $r<1$.

By lemma \ref{corresp2lemma}, the vector $\vec v_a$ is in one-to-one correspondence with a tuple $(\vec x,f_1,\ldots,f_{a-1})$ where $\vec x$ is a vector in lattice $\Gamma_0^{i_a}/\tilde K^P_{a-1}$.
By lemma \ref{diamboundlemma}, the lattice $\Gamma_0^{i_a}/\tilde K^P_{a-1}$ has the diameter of its Voronoi cells bounded by $\sqrt{i_a-a+1}$.
So, for given $\Gamma_0^{i_a}/\tilde K^P_{a-1}$ and given $r$, the number of possible choices of $\vec x$ such $|\pi_{a-1}(\vec v_a)| \leq r$
is bounded by
\be
N(\Gamma_0^{i_a}/\tilde K^P_{a-1},0,r) \leq \frac{1}{\vol(\Gamma_0^{i_a}/\tilde K^P_{a-1})} V_{i_a-a+1}(r+\sqrt{i_a-a+1}).
\ee
So, by Eq.~(\ref{volinvert}),
\be
N(\Gamma_0^{i_a}/\tilde K^P_{a-1},0,r) \leq \vol(K^P_{a-1}) V_{i_a-a+1}(r+\sqrt{i_a-a+1}).
\ee

Factorize 
$M_K(i_a,a-1)=M_{{\tilde K^P}_{a-1}} F$, as in lemma \ref{corresp2lemma}.
 
The number of possible choices of $f_1,\ldots,f_{a-1}$ is equal to ${\rm det}(F)=\vol(K_{a-1})/\vol(K^P_{a-1})$.

So, the total number of choices of $K_a$ is bounded by
\be
{\rm det}(F) \vol(K^P_{a-1}) V_{i_a-a+1}(r+\sqrt{i_a-a+1})=\vol(K_{a-1})  V_{i_a-a+1}(r+\sqrt{i_a-a+1}),
\ee
as claimed.
\end{proof}
\end{lemma}

\subsection{First Moment Bound}
\begin{lemma}
\label{fmb}
Let $G$ be an $n$-by-$k$ code generator matrix for a code, chosen from the ensemble defined previously (entries chosen independently and uniformly from $\mF_p$).
Let $M_K$ be an $n$-by-$k$ lattice generator matrix.
Let $K_{a-1}$ be given and assume the first $a-1$ columns of $M_K$ are in Hermite normal form.
Let $Pr(K_{a-1},r)$ denote the probability that, conditioned on $K_{a-1}$ being consistent with $G$, there exists a choice of $v_a$ such that 
$K_a$ is consistent with $G$ and such that the first $a$ columns of $M_K$ are in Hermite normal form
and such that 
\be
\vol(K_{a}) \leq r K_{a-1}.
\ee
Then, for $r<1$, $Pr(K_{a-1},r)=0$, and for $r< p$,
\be
Pr(K_{a-1},r) \leq p^{-(n-k)} \vol(K_{a-1})  V_{i_a-a+1}(r+\sqrt{i_a-a+1}).
\ee
\begin{proof}
By lemma \ref{colcountlemma}, indeed there are no choices of $v_a$ such that $r<1$.
If $r< p$, then $0<(\vec v_a)_1<p$ so $(\vec v_a)_1 \neq 0 \mod p$.
So, the $a$-th column of $M_K$ is not in the span of the first $a-1$ columns of $M_K$ modulo $p$.
So, even though we have conditioned on $K_{a-1}$ being consistent with $G$, the probability that a given choice of $\vec v_a$ is consistent with  $G$
is bounded by $p^{-(n-k)}$.  (The probability is $p^{-(n-k)}$ if we condition on $G$ being non-degenerate and smaller if $G$ may be degenerate.)

So, by lemma \ref{colcountlemma}, the average number of choices of $\vec v_k$ consistent with $G$ is bounded by
$p^{-(n-k)} \vol(K_{a-1})  V_{i_a-a+1}(r+\sqrt{i_a-a+1})$.
\end{proof}
\end{lemma}

The next theorem estimates the probability that, for a randomly chosen code generator matrix, there is a rank-$m$ lattice $K$ of small volume which is consistent with that
matrix.  The bounds becomes effective for volume smaller than $(cp)^{{\rm min}(m,n-k)}$ with $c<1/\sqrt{2\pi e}$.
\begin{theorem}
\label{countingthm}
 Let $P_{lat}(H,p,n,m)$ denote the probability that for a random code generator matrix $G$ for a code over $\mF_p^n$ there is a rank-$m$ lattice $K$ consistent with a code generator matrix such that $\vol(K) \leq H$.

For any $p$, for any real number $x>\sqrt{2\pi e}$, for sufficiently large $n-m$, 
\be
P_{lat}((cp)^{{\rm min}(m,n-k)},p,n,m) \leq m  c^{m} x^{n-m+1}.
\ee
The required $n-m$ is quadratic in $p(x-\sqrt{2\pi e})^{-1}$.
\begin{proof}
Note that if there is a lattice $K$ of rank-$m$ consistent with the code generator matrix, then the lattices $K_1,\ldots,K_{m-1}$ constructed above have ranks $1,\ldots,m-1$ respectively and are also
consistent with the code generator matrix and have $\vol(K_a) \leq \vol(K)$.
So, it suffices to consider the case $m\leq n-k$ (if $m>n-k$, then consider the lattice $K_{n-k}$ instead).

For $M_K$ in Hermite normal form, since $i_1<i_2<\ldots<i_m<n$, we have  $i_a \leq n-m+a$ and so $i_a-a+1 \leq n-m+1$.
We use the bound (the inequality on the second line holds for all sufficiently large $n-m$)
\begin{eqnarray}
\label{vineq}
V_{n-m+1}(r+\sqrt{n-m+1})&=&\frac{\pi^{\frac{n-m+1}{2}}}{\Gamma(\frac{n-m+1}{2})}  (r+\sqrt{n-m+1})^{n-m+1} \\ \nonumber
&\leq &\Bigl( \frac{2\pi e}{n-m+1}\Bigr)^{\frac{n-m+1}{2}} (r+\sqrt{n-m+1})^{n-m+1} \\ \nonumber
&=& \Bigl(\frac{r\sqrt{2\pi e}}{\sqrt{n-m+1}}+\sqrt{2 \pi e} \Bigr)^{n-m+1},
\end{eqnarray}

Let $c$ be a real number, $0<c<1$.  We will make a choice of $c$ below.

By lemma \ref{fmb} and Eq.~(\ref{vineq}), given $K_{a-1}$, if $\vol(K_{a-1}) \leq (cp)^m$, 
we have
\begin{eqnarray}
Pr(K_{a-1},r) & \leq & c^m p^{m-(n-k)}  \Bigl(\frac{r\sqrt{2\pi e}}{\sqrt{n-m+1}}+\sqrt{2 \pi e} \Bigr)^{n-m+1} \\ \nonumber
& \leq & c^m  \Bigl(\frac{r\sqrt{2\pi e}}{\sqrt{n-m+1}}+\sqrt{2 \pi e} \Bigr)^{n-m+1}.
\end{eqnarray}
For $r<p$, this is bounded by $c^{m}  \Bigl(\frac{p \sqrt{2\pi e}}{\sqrt{n-m+1}}+\sqrt{2 \pi e} \Bigr)^{n-m+1}$.
For any $p$, for any real number $x>\sqrt{2\pi e}$, for sufficiently large $n-m$, this is bounded by $c^{m} x^{n-m+1}$. (The required $n-m$ is quadratic in $p(x-\sqrt{2\pi e})^{-1}$).

Suppose that $\vol(K) \leq (cp)^{m}$ for some $c<1$.
Then, $\vol(K_a)\leq (cp)^{m}$ for all $a$ and for some $a$ we have $\vol(K_a)/\vol(K_{a-1})<p$.
However, for $\vol(K_a) \leq (cp)^{m}$, the above calculation bounds the probability for given $a$ that there is a choice of $K_a$ such that $\vol(K_a)/\vol(K_{a-1})<p$ by $c^m x^{n-m+1}$ for all sufficiently large $n-m$.  By a union bound, the probability that for some $a$ there is
a choice of $K_a$ such that $\vol(K_a)/\vol(K_{a-1})<p$ is bounded by $m c^m x^{n-m+1}$ for all sufficiently large $n-m$.
So, $P_{lat}((cp)^{m},p,n,m) \leq m c^m x^{n-m+1}$ for all sufficiently large $n-m$.
\end{proof}
\end{theorem}

This implies the following corollary for the Rankin invariant:
\begin{corollary}
\label{rankincor}
For any $p,k$, for all sufficiently large $n$ at fixed ratio $m/n$, for any $c<1/\sqrt{2\pi e}$,
with high probability
we have
\be
\gamma_{n,m}(L_0)\geq (cp)^{2{\rm min}(m,n-k)} p^{-2m(n-k)/n}.
\ee
\end{corollary}
(Recall that with high probability $G$ is non-degenerate so $L_0$ is rank $n$.)

We remark that the bounds of theorem \ref{countingthm}, the bounds on the constant $x$ may not be tight, especially for small $m$.
One possible way to tighten the bounds is to use the fact that if there $\vol(K)<p^{m-z}$ for some integer $z>0$ then there must be at least $z$ different $a$ such that $\vol(K_a)/\vol(K_{a-1})<p$; in the proof above we only used that there was at least one such $a$.

We remark also that, up to the constant $c$, the value of the Rankin invariant at $m=k=n/2$ is optimal for an integer lattice; i.e., the dependence on $p$ is optimal.  The reason is that it implies that an $n/2$-dimensional sublattice of $L_0$ has the same volume (again, up to factors of $c^{m}$) as $L_0$ does.

It is also worth comparing the value of the Rankin invariant that we find to the Rankin invariant for random lattices (from a different ensemble) in Ref.~\onlinecite{blockwise}.  
The Rankin constant $\gamma_{n,m}$ is defined to be the maximum of $\gamma_{n,m}(L)$ over all lattices $L$.
Those random lattices in Ref.~\onlinecite{blockwise} were used to lower bound the Rankin constant $\gamma_{n,n/2}$ by $\gamma_{n,n/2} \geq (\frac{k}{12})^{n/4}$.  Since we need to take $n \sim p^2$ for the bounds of theorem \ref{countingthm} to be effective, if we choose $m=k=n/2$ and $p\sim \sqrt{n}$ we find that with high probability $\gamma_{n,n/2}(L_0) \geq ({\rm const.} \times n)^{n/4}$.  Thus, we find the same leading behavior $n^{n/4}$, with the Rankin invariants differing only by factors ${\rm const.}^{n}$.

\section{Volume of Oriented Systole}
\label{calibrationsec}
In this section, we consider a weaker conjecture than \ref{conj1}.  Throughout this section, we consider the case of homology using integer coefficients, rather than $\mZ_2$ coefficients.
In this setting, there is a general method, called ``calibration"\cite{calibration} for lower bounding weights.
We will show that this method gives an effective lower bound for homology class eswhich have a particular form, which we call ``split", but we will
show that it does not give a useful lower bound in general.  The reason for this is related to the existence of short vectors in the exterior $q$-th power of $L_0$.

Given an rank-$n$ lattice $L$, we write its $m$-th exterior power as $\wedge^m L$.  This exterior power is a lattice of vectors in
${n \choose m}$ dimensions; the vectors in this lattice are linear combinations (with integer coefficients) of vectors $v_1 \wedge v_2 \wedge \ldots \wedge v_m$,
where $v_i \in L$ and the exterior product is anti-symmetric under interchange: $v_1 \wedge v_2 = - v_2 \wedge v_1$.
\begin{definition}
A vector $v$ in $\wedge^m L$ is called ``split" if $v=x_1 \wedge \ldots \wedge x_m$ for $x_1,\ldots x_m \in L$.
\end{definition}

The $q$-th homology classes of the torus $T^n$ are in one-to-one correspondence with vectors in $\wedge^q {\mathbb Z}^n$.
For the torus ${\mathbb R}^n/L_0$ that we consider, it will be more convenient to regard the classes as being in one-to-one correspondence with vectors in $\wedge^q L_0$.
That is, the $k$-th homology class represented by a hyperplane which is a span of $k$ basis vectors will correspond to the vector which is the exterior product of these $k$ basis vectors.

The lattice $\wedge^m L$ inherits an inner product:
$$(x_1 \wedge \ldots \wedge x_m) \cdot (y_1 \wedge \ldots \wedge y_m)={\rm det}(S),$$
where $S$ has matrix elements $S_{i,j}=x_i \cdot y_j$.
We write this norm $|X|$, where $X\in \wedge^m L_0$.
Calibration allows one to lower bound the volume of a representative of a homology class
in $\wedge^q L_0$ using this inner product.

We first explain this lower bound in the split case.  The arguments are not new.
\begin{lemma}
\label{calsplit}
Let $\vol(v_1,\ldots,v_q)\neq 0$.  Then, the minimum volume of any closed chain (either a sum of faces of $q$-faces of the unit hypercubes used in the cubulation or more generally an arbitrary sum of simplices) representing homology class
$X=v_1 \wedge \ldots \wedge v_q$ is greater than or equal to $|v_1 \wedge \ldots \wedge v_q|$.
\begin{proof}
Let us write $v \cdot d \vec x$ to denote a differential $1$-form $\sum_i (v)_i d x^i$, where $i=1,\ldots,n$ are orthogonal basis directions in Euclidean space and $(v)_i$ are components of $i$.
Consider the differerential $q$-form $\omega=(v_1 \cdot d\vec x) \wedge (v_2 \cdot d \vec x) \wedge \ldots \wedge (v_q \cdot d \vec x)$.
Let $S$ denote the hyperplane spanned by vectors $v_1,\ldots,v_q$ (the hyperplane is oriented, so the order of vectors matters).
We have $\int_S \omega=|X|^2$.
Further, for any chain $C$ in the same homology class as $S$, we have $\int_C \omega=\int_S \omega=|X|^2$, where
the integral over $C$ is given by writing $C$ as a sum of $q$-faces of the unit hypercubes and integrating $\omega$ over each face.
(Indeed, one can also consider more general $C$, such as sums of arbitrary simplices, and the same result holds).
For a $q$-face (or indeed any sum of $q$-dimensional simplices), the integral of $\omega$ over that face is bounded by
$|X|$ times the volume of the face.
Hence, the volume of $C$ must be at least equal to $(\int_C \omega)/|X|=|X|$.
\end{proof}
\end{lemma}

Now we consider the nonsplit case.  In contrast to the split case where we were able to ``calibrate" the hyperplane $S$ (find a differential form assuming maximum value on that hyperplane), we might not be able to calibrate nonsplit homology classes.
However,
we can still obtain a lower bound.
\begin{lemma}
\label{calgen}
Let $X\in \wedge^q L$, $X \neq 0$.
Then, the minimum volume of any closed chain representing homology class $X$ is lower bounded by $|X|$.
\begin{proof}
Write $X=\sum_a X_a$, where $X_a$ are split vectors.
For each $X_a=v_1^a \wedge \ldots \wedge v_q^a$, define a differential $q$-form
$\omega_a=(v_1^a \cdot d\vec x) \wedge \ldots \wedge (v_q^a \cdot d \vec x)$.
Let $\omega=\sum_a \omega_a$.

Let $S_a$ denote the hyperplane spanned by vectors $v^a_1,\ldots,v_q^a$.
Let $S$ denote the union of hyperplanes $S_a$.
We have $\int_{S_a} \omega_b=(X_a,X_b)$.
Hence, $\int_S \omega=|X|^2$.

We now consider the maximum of the integral of $C$ over a $q$-face or $q$-dimensional simplex of unit volume.
This is equal to $${\rm max}_{V \; {\rm split}, \; |V|=1} (V,X),$$ where we take the maximum over all split vectors $V\in \wedge^q {\mathbb R}^n$, with $V$
not necessarily in $\wedge ^q L$; i.e., $V=v_1 \wedge \ldots \wedge v_q$ for {\it arbitrary} $v_1,\ldots v_q$, with $v_1,\ldots, v_q$ not necessarily in the lattice $L$ (i.e., we are upper bounding the integral over a unit volume square in the hyperplane spanned by $v_1,\ldots v_q$).
If we relax the requirement that $V$ be split, we have ${\rm max}_{V} (V,X)=|X|$.  The restriction
to split $V$ can only reduce the maximum, so the maximum over split $V$ is at most $|X|$.
So, as in lemma \ref{calsplit}, since the integral over $\omega$ over any chain representing the same
homology class as $X$ must be equal to the $\int_S \omega=|X|$, the volume of such a chain must be at least $|X|$.
\end{proof}
\end{lemma}

One may wonder whether the bound in lemma \ref{calgen} can be significantly improved if we do not relax the requirement that $V$ be split.  Of course, if $X$ is split, then ${\rm max}_{V \; {\rm split}, \; |V|=1} (V,X) \geq |X|/\sqrt{{n \choose q}}=|X|$ and the maximum is
achieved for $V=X$.  However, for $X$ not split, the maximum might be smaller and so the lower bound on the volume would be correspondingly: we can lower bound the volume of a closed chain representing homology class $X$
by $|X|^2/{\rm max}_{V \; {\rm split}, \; |V|=1} (V,X)$.
Unfortunately, this at best only leads to a small improvement in the bound.  We claim that
\be
\label{splitV}
{\rm max}_{V \; {\rm split}, \; |V|=1} (V,X) \geq |X|/\sqrt{{n \choose q}},
\ee
so that at best we would lower bound the volume by $\sqrt{{n \choose q}} |X|$, and since $\sqrt{{n \choose q}}< 2^{n/2}$, this leads to only
a small improvement (recall that there are $N=p^{n/2}$ qubits and we choose $p>>1$).
To see Eq.~\ref{splitV}, consider the orthogonal basis for $\wedge^q {\mathbb R}^n$ of vectors $x_1 \wedge \ldots \wedge x_q$ where $x_1,\ldots, x_q$ are chosen from the $n$ different coordinate directions.  These basis vectors are all split.  Since $\wedge^q {\mathbb R}^n$ is ${n \choose q}$-dimensional, there must be some basis vector $V$ such that $|(V,X)| \geq |X|/\sqrt{{n \choose q}}$.  Using this vector $V$ (or its negation if the inner product $(V,X)$ is negative) in the maximum gives Eq.~(\ref{splitV}).

The Rankin invariant is the minimal value of the norm $|X|$ over nonzero split vectors.  Thus, the results on the Rankin
invariant give a lower bound on the volume of representatives of split homology classes.
However, in Ref.~\onlinecite{coulangeon}, it was shown that for certain lattices $L$ the shortest nonzero vector in $\wedge^m L$ may be shorter than the Rankin invariant.
Interestingly, the lattices we consider here provide another example where this occurs; in fact this occurs for any lattice with sufficiently large Rankin invariant.
\begin{lemma}
\label{shortsplit}
Let $L$ be a rank-$n$ lattice.
Then, the shortest  nonzero vector in $\wedge^m L$ has norm at most 
$\sqrt{\vphantom{I} \gamma_{{n \choose m}} }\vol(L)^{m/n}$, where
$\gamma_{{n \choose m}}$ denotes Hermite's constant in dimension ${n \choose m}$.

Hence, if $\gamma_{n,m}(L) \geq \gamma_{{n \choose m}}$, then the shortest vector is not split.
\begin{proof}
We have $\vol(\wedge^m L)=\vol(L)^{{n-1} \choose {m-1}}$ by Proposition 1.10.4 of Ref.~\onlinecite{perfect}.
The lattice $\wedge^m L$ has rank $r={n \choose m}$, and so
the shortest nonzero vector in $\wedge^m L$ has length at most
$\sqrt{\gamma_{r}} \vol(\wedge^m L)^{1/r}$, where $\gamma_r$ is Hermite's constant.
So, the shortest nonzero vector in $\wedge^m L$ has length at most
\be
\sqrt{\gamma_r} \vol(L)^{{{n-1} \choose {m-1}}/{n \choose m}}=\sqrt{\gamma_r} \vol(L)^{m/n}.
\ee
\end{proof}
\end{lemma}

For all $r$, we have $\gamma_r\leq 1+r/4$, with an asymptotic behavior $\gamma_r \lesssim \frac{2r}{\pi e}$\cite{MH}.
So, $\sqrt{\vphantom{I}\gamma_{{n \choose m}}} \leq \sqrt{1+{n \choose m}/4}$.
So, lemma \ref{shortsplit} has an interesting interpretation for the application to quantum codes.
If the bound in lemma \ref{calgen} is saturated so that the least volume cycle representing a homology class has volume $|X|$, then we find that the code has roughly square-root distance.
Thus, conjecture \ref{conj1} implies that for some homology classes, the bound of lemma \ref{calgen} is far from saturated.
The possible improvement of Eq.~(\ref{splitV}) leads to only a small improvement here (though, it is possible that if the possible improvement of Eq.~(\ref{splitV}) holds for the homology classes with smallest $|X|$ and if the bound of lemma \ref{shortsplit} is saturated then one might be able to prove a slightly above square-root distance for integer homology).

\section{Quantum Locally Testable Codes from High-Dimensional Constructions}
\label{qltcsec}

In this section, we give a construction of quantum codes which are ``locally testable"\cite{qltc} using high-dimensional constructions.
The construction uses a different topology than above; the similarity in the constructions is simply that in both cases we consider a family of codes derived from manifolds of varying dimension, with the number of qubits in the code depending exponentially on the dimension of the manifold.

Let us write $\wt(O)$ to indicate the weight of an operator $O$.  Similarly, given a vector $v$ (in one of the vector spaces defining the chain complexes), we let $\wt(v)$ denote the number of nonzero entries in $v$.

Given a  CSS stabilizer code defined from a chain complex $\ldots \mC_{q+1} \stackrel{\partial_{q+1}}{\rightarrow} \mC_q \stackrel{\partial_q}{\rightarrow} \mC_{q-1} \ldots$, with the qudits associated with $q$-cells and the $Z$-type and $X$-type stabilizers associated with $(q+1)$-cells and $(q-1)$-cells, respectively,
we define soundness parameters $\epsilon_X(w),\epsilon_Z(w)$ as follows:
\begin{definition}
Define
\be
\epsilon_Z(w)={\rm min}_{v\in \mC_q,\wt(v)=w, \partial_q v \neq 0} \Bigl( {\rm max}_{u \in \mC_q, \partial_q u=0} \frac{\wt(\partial v)}{\wt(v+u)}\Bigr).
\ee
Define $\epsilon_X(w)$ similarly, with $\partial_q$ replaced with $\partial_{q+1}^T$, where the superscript $T$ denotes transpose.
\end{definition}
Equivalently, consider the minimum over all $Z$-type operators $O$, such that $O$ has weight $w$ and such that $O$ does not commute with at
least one stabilizer, of the following quantity: take the maximum, over all $Z$-type operators $P$ which commute with all stabilizers, of the ratio of the number of stabilizers which do not commute with $O$ to the weight of $O+P$.  This minimum is $\epsilon_Z$.

It is unclear whether or not families of codes exist which have distance which is $\Omega(1)$ and stabilizer weight $\mO(1)$ and
which have $\epsilon_{X,Z}(w)$ bounded away from zero for all $w$.  However, the codes of Ref.~\onlinecite{tz} have
distance $\Theta(\sqrt{N})$, stabilizer weight $\mO(1)$ and have $\epsilon_{X,Z}(w)$ bounded away from zero for $w\lesssim \sqrt{N}$, as shown in
Ref.~\onlinecite{tzqltc}

Here we give a simple construction of a family of qubit codes with $2$ encoded qubits and with distance $\Theta(\sqrt{N})$, $\epsilon_{X,Z}(w)$ only polylogarithmically small for all $w$, and with {\it logarithmic} weight stabilizers.
We warm up with a construction of a qubit code family with no encoded qubits (and hence the notion of distance is meaningless for this code) but with $\epsilon_{X,Z}(w)$ bounded away from zero for all $w$ and with logarithmic weight stabilizers; we call this the ``simplex code".  We then give the full construction, which is based on a product of hyperspheres.

\subsection{Simplex Code}
Of course, with no encoded qubits, there are some fairly trivial constructions of code with $\epsilon_{X,Z}$ strictly bounded away from zero.  For example, one can take a code with $N$ qubits and stabilizers $Z_1,Z_2,\ldots,Z_N$.  Thus, every product of $Z$ operators commutes with all stabilizers (and so $\epsilon_Z(w)$ is a minimum over an empty set), while clearly $\epsilon_X(w)=1$ for all $w$.
However, the simplex code construction that we give obeys Poincare duality and has an entangled ground state.

The code we consider is obtained by taking a toric code on a $n$-dimensional sphere, with the degrees of freedom on $q$-cells for $q =n/2$.  The exact value of $q$ is not very important; the important thing is that $q/n$ is neither close to $0$ nor close to $1$ so that the number of $r$-cells will be exponential in $n$.  However, the case $q=n/2$ is the self-dual case so this makes the proofs slightly simpler as we need to consider only one type of stabilizer.

The cellulation of the $n$-sphere that we use is to take the boundary of a $n+1$-dimensional simplex.
We label the $0$-cells by integers $1,\ldots,n+2$.  For $0 \leq r \leq n$, there are ${n+2 \choose r+1}$ distinct $r$-cells, labelled by subsets of $\Lambda\equiv \{1,\ldots,n+2\}$ with $r+1$ elements.
We use qubits so the vector spaces are all over $\mF_2$.

For $1 \leq r \leq n$, the boundary operator $\partial_r$ acting on an $r$-cell labelled by some $(r+1)$-element set $S \subset \Lambda$ gives the sum of $r+1$ different $(r-1)$-cells, labelled by the distinct $r$-element subsets of $S$.
For example, for $n\geq 2$, $\partial_2 \{1,2,3\}=\{1,2\}+\{1,3\}+\{2,3\}$.
We set $\partial_0=0$.  One may verify that $\partial_{r-1} \partial_r=0$ for all $r$.

For $q=n/2$, there are $N={n+2 \choose n/2+1}$ qubits, so $N$ is exponentially large in $n$.  Remark: in previous sections, the number of qubits we also had
an exponential factor $p^{n-k}$ which, for large $p$, was the dominant exponential scaling; in this subsection, we do not have such a factor.

Each qubit is acted on by $q+1$ stabilizers (as each $q$-cell has $q+1$ cells in its boundary)
and each stabilizer acts on $q+2$ different qubits (as each $(q+1)$-cell has $q+2$ cells in its boundary and each $(q-1)$-cell has $q+2$ cells in its coboundary).
Hence, the weight is indeed logarithmic in $N$, $w=(1/2+o(1)) \cdot \log_2(N)$.

Finally, we show soundness.  First, let us introduce notation.
\begin{definition}
Given an $r$-cell $\sigma$ labelled by some set $S$ and a set $T\subset \Lambda$,
we define $r \cup T$ to equal $0$ if $S\cap T\neq \emptyset$ and
otherwise $r \cup T$ is
the $r+|T|$-cell labelled by $S \cup T$.

Given a vector $v \in \mC_r$, we define $v \cup T$ by linearity.  $v \cup T \in \mC_{r+|T|}$ and
the coefficient of $v \cup S$ corresponding to an $r+|T|$-cell labelled by a set $U$ is equal to the
coefficient of $v$ corresponding to the $r$-cell labelled by $U\setminus T$ if $T \subset U$ and is equal to
$0$ is $T \not\subset U$.
\end{definition}

\begin{lemma}
For the simplex code, for all $w$,
$\epsilon_X(w)=\epsilon_Z(w) \geq 1$.
\begin{proof}
Consider any $v\in \mC_q$ with $\partial_q v \neq 0$.
Set $w=(\partial_q v) \cup \{1\}$.
Then, one may verify that $\partial_q x = \partial_q v$ (and hence, setting $w=x-v$, $\partial_q w=0$) and that
$\wt(x) \leq \wt(\partial_q v)$.
\end{proof}
\end{lemma}

The proof of soundness above has a very simple geometric interpretation.  We take the boundary $\partial_q v$ and shrink it to a point (arbitrarily choosing the vertex $\{1\}$ as the point that we shrink it to).

\subsection{Hypersphere Product Code}
The above construction had constant soundness, but had no encoded qubits.  We now give a different construction with $2$ encoded qubits and
distance $\sqrt{N}$ and inverse polylogarithmic soundness. We now consider the toric code on a product of spheres, $S^n \times S^n$.

We pick an integer $p \geq 1$ ($p$ need not be prime); $p$ will be chosen to equal $\log(N)$ below in order to
achieve square-root distance. 
We choose a cellulation of $S^n$ as follows: consider an $(n+1)$-dimensional hypercube of side length $p$ on each side (we call this the ``large" hyercube).  Cellulate that large hypercube using hypercubes of side length $1$ on each side; we call these the ``small" hypercubes) (so that there are $p^{n+1}$ small hypercubes in the cellulation).  Then, take the boundary of the hypercube to get a cellulation of $S^n$.

A small $(n+1)$-dimensional hypercube has ${n+1 \choose r} 2^{n+1-r}$ different $r$-cells in its boundary (each $r$-cell is a product of $1$-cells in $r$ out of the $n+1$ directions and then for each of the remaining directions there are $2$ possible choices of $0$-cells).
The number of $r$-cells in the cellulation of the large hypercube is ${n+1 \choose r} (p+1)^{n+1-r} p^r$.  To see this, assign coordinates $[0,p]$ for each side of the large hypercube.  Then, each $r$-cell is a product of $1$-cells in $r$ out of the $n+1$ directions with $0$-cells in the remaining directions.  The midpoints of the $1$-cells are at half-integer coordinate in the interval $[0,p]$ and so there are $p$ possible choices for each cell.  There are $p+1$ possible choices for the coordinates of each $0$-cell as these cells are at integer coordinates in the interval $[0,p]$.
To determine the number of $r$-cells in the boundary of the large hypercube, restrict to the case that in at least one of the directions, the coordinate must be $0$ or $p$.  This gives the number equal to
$${n+1 \choose r} (p+1)^{n+1-r} p^r \Bigl(1-(\frac{p+1-2}{p+1})^{n+1-r}\Bigr),$$
where
the ratio in parenthesis $\frac{p+1-2}{p+1}$ is the probability that for a random integer coordinate in the range $[0,p]$, the coordinate is not on the boundary $0$ or $p$.
Thus, there are at most $2^{(1-o(1))\cdot n} p^n$ cells (if $n>>p$) and at least $2^{1-o(1))\cdot n} p^{n-1}$ cells (if $n<<p$).

We take the product of this cellulation with itself to get a cellulation of $S^n \times S^n$.
The degrees of freedom will be on the $q$-cells for $q=n$, so that $N$ is again exponential in $n$.  We have
$\log_2(N) =2(1+\log_2(p)+o(1))\cdot n$ for $n>>p$ and $\log_2(N)=2(1+\log_2(p)+o(1))\cdot n - 2\log_2(p)$ for $n<<p$.
We will take $p=1/\log(N),n=\Theta(\log(N)/\log(\log(N)))$.

Each qubit is acted on by $2n$ stabilizers and each stabilizer acts on $2(n+1)$ qubits.  Hence, the weight is logarithmic in $N$, $w=\Theta(\log(N)/\log(\log(N)))$.

The number of encoded qubits is equal to $2$, as can be computed from the homology of $S^n \times S^n$ (by the K\"{u}nneth formula, $H_i(S^n \times S^n;\mZ_2)=2$ for $i=n$, 
$H_i(S^n \times S^n;\mZ_2)=1$ for $i=0,2n$, and $H_n(S^n \times S^n;\mZ_2)=0$ otherwise).

\begin{lemma}
For the hypersphere product code,
\be
d_X(w)=d_Z(w)=(p+1)^{n+1}\Bigl(1-(\frac{p+1-2}{p+1})^{n+1}\Bigr)=\Theta(N^{\frac{p}{p+2}}).
\ee
For $p=\Omega(\log(N))$,
\be
d_X(w)=d_Z(w)=\Theta(\sqrt{N})).
\ee
\begin{proof}
Let $a$ be a $0$-cell in the second $S^n$ in the product $S^n \times S^n$.
Let
$Z(a,2)$ be the logical $Z$ operator which is the product
 $Z_i$ over all $i$ which are the product of an $n$-cell in the first $S^n$ with $0$-cell $a$.
 Then, any logical $X$ operator which anticommites with $Z(a,2)$ must have some support on some cell $i$ which is a product of an $n$-cell in the first $S^n$ with $0$-cell $a$.  However, since $Z(a,2)$ and $Z(b,2)$ are homologous for any two choices of $0$-cells $a,b$ in the second $S^n$ ($Z(a,2),Z(b,2)$ differ by a product of stabilizers), that logical $X$ operator must have some support 
on some cell $i$ which is a product of an $n$-cell in the first $S^n$ with $0$-cell $a$ for {\it all} $0$-cells $a$ in the second $S^n$.
Hence, that logical $X$ operator must have a number of cells in its support equal to the number of $0$-cells in the second $S^n$.
This number is equal to 
$(p+1)^{n+1}\Bigl(1-(\frac{p+1-2}{p+1})^{n+1}\Bigr).$

This number is also an upper bound to $d_X(w)$, since the product of $X$ over all cells which are a product of
a fixed $n$-cell in the first $S^n$ with an arbitrary $0$-cells in the second $S^n$ is a logical operator.

We can similarly lower bound the number of cells in the support of any logical $X$ operator which anticommutes with
the operator $Z(a,1)$, defined to be the logical $Z$ operator which is the product
 $Z_i$ over all $i$ which are the product of an $n$-cell in the {\it second} $S^n$ with $0$-cell $a$ in the {\it first} $S^n$.
\end{proof}
\end{lemma}

We now show soundness.  Again, the geometric interpretation is to shrink the boundary to a point.
\begin{lemma}
For the hypersphere product code, $\epsilon_X(w)=\epsilon_Z(w) \geq \Omega(1/\log(N)^2)$.
\begin{proof}
Consider any $v\in \mC_q$ with $\partial_q v \neq 0$.

We place coordinates $[0,p]^{n+1}$ on the first large hypercube.  Call the face where the first coordinate is equal to $p$ the ``top face".
Call the face where the first coordinate is equal to $0$ the ``bottom face". 
Let $v_0=v$.  We will construct a sequence
$v_1,v_2,\ldots, v_f\in \mC_q$ for some integer $f$ where we bound $\wt(v_{i+1}-v_i)$ with the final vector $v_f=0$.
In this way, we will bound $\wt(v_0)$.
We construct the sequence so that the boundaries $\partial_q v_i$ are first removed from the top face of the first hypercube, then moved from the top face to the bottom face of the first hypercube, and finally moved to a point on the bottom face of the first hypercube.

Throughout this proof, when we refer to coordinates, we refer to the first hypercube in the product.
We regard an $r$-cell as being a product of $0$-cells and $1$-cells.  That is, each cell in the product of hypercubes is product of cells in each hypercube.  Then, each $r$-cell in a hypercube is a product of $r$ $1$-cells and $n+1-r$ $0$-cells.  The $n+1$ different
terms in the product correspond to different coordinates.
When we say that a cell ``is a $0$-cell" in a given coordinate, we mean that the cell is a product of a $0$-cell in that coordinate with
some cells in other coordinates.

We first explain the middle step, moving from top face to bottom face.  Suppose that some $v_i$ has $\partial_q v_i$ vanishing on the top face.
Indeed, suppose that $\partial_q v_i$ vanishes if the first coordinate is greater than $x$, for some integer $x$.
Then, let $\pi_x(\partial_q v_i)$ be the projection of $\partial_q v_i$ onto cells with first coordinate equal to $x$.
This projection consists only of cells which are $0$-cells in the first coordinate.  Let $v_{i+1}-v_i$ be defined by taking $\pi_x(\partial_q i)$ and
replacing every $0$-cell in the first coordinate at position $x$ with a $1$-cell at position $x-1/2$.
Then, $\pi_x(\partial_q v_{i+1})=0$.  Iterating this procedure, decreasing $x$ from $p$, to $p-1$, to $p-2$, and so on, we can construct a sequence $v_i,v_{i+1},\ldots$ so that the final vector in the second has boundary only on the bottom face.  There are at most $p$ steps in the sequence.  Note that because $\partial^2=0$, once we ensure that $\pi_x(\partial_q v_i)=0$, then we know that
$\partial_q v_i$ has not cells which are $1$-cell in the first coordinate with midpoint as $x-1/2$.

Now we explain the first step, moving the boundary off the top face.  We apply above the above procedure to the {\it second} coordinate.  We let $\pi_{p,x}(\partial_q v_i)$ be the projection $\partial_q v_i$ onto cells with first coordinate equal to $p$ and second coordinate equal to $x$, for integer $x$.  We then construct a sequence so that this projection vanishes for $x=p,p-1,\ldots$, following the same procedure as in the above paragraph.  There are at most $p$ steps in this sequence.  We then repeat this for the second coordinate, third coordinate, and so on; giving at most
$pd$ steps.

The final step is the same as the first step, with the top face replaced by the bottom face.

So, there are at most $\mO(pd)$ steps in the sequence.
We have $\wt(\partial_q v_i) \leq \wt(\partial_q v)$ for all vectors in the sequence, and there are at most $\mO(pd)$ steps, so
this gives $\epsilon_X(w)\geq \Omega(1/pd)=\Omega(1/\log(N)^2)$.
\end{proof}
\end{lemma}

In the above construction, we lost a factor of $d$ due to having $d$ steps in the sequence to move the boundary.  Likely for this
construction, this factor cannot be avoided since the diameter of the hypercube is $pd$.  One might wonder whether other geometries (such as a geometry that more closely approximates a sphere) would improve on this factor; note however that since the volume of a sphere of radius $r$ in $d$-dimensional Euclidean space scales roughly as $(r/d)^{d/2}$, one would need to take the radius proportional to $d$ in order to obtain a large volume so again one would need to have a large diameter for the geometry.

\section{Discussion}
We have presented several different code constructions based on the toric code on families of higher-dimensional manifolds.  Rather than varying the geometry or topology at fixed dimension, as is more commonly done, we have considered varying dimension.  This leads to a scaling in which the number of qubits, $N$, scales exponentially with dimension, $n$, so that the weight of the stabilizers $w$ is proportional $n \propto \log(N)$.  Assuming conjecture \ref{conj1}, we have constructed a code family with almost linear distance and logarithmic weight.

{\it Acknowledgments---} I thank L. Eldar and M. Freedman for useful discussions.

\end{document}